\documentclass[12pt]{article}
\usepackage{blois,graphicx}

\begin{document}

\title{Baryon String-Junction Torus Exchange: an Appearance in High Energy Proton Collisions}

\author{Olga I. Piskounova}

%
\address{P.N. Lebedev Physical Institute of Rusiian Academy of Sciences, Leninski prosp. 53, 119991 Moscow, Russia}
\date{Received: 25.09.2019 / Revised version: 25.09.2019}
%
\maketitle\abstract{The topological presentation of pomeron exchange at the proton-proton collision of high energy is cylinder that is covered with the net of quark-gluon exchanges. I suggest that the process of double diffraction dissociation (DD) can be presented as one pomeron exchange with the central loop of two uncut pomeron cylinders.  Taking into account that the junction of three gluon strings (SJ) has the positive baryon number, as well as the antijunction is of negative baryon charge, our neutral pomeron construction can be covered with only a certain number of hexagons that are built of 3 junction and 3 antijunction each. This image is similar somehow to graphene tube. It is reasonable to expect that the dynamics of rapidity gaps in DD should be determined by the number of hexagons on the surface of pomeron torus. Therefore, the gap distribution in DD events has the discrete structure in the region of large gaps. Moreover, the SJ torus can be released in pp interactions as metastable particle and is getting suspected as Dark Matter candidate. The possibility of production of multi-quark states with few string junctions has been discussed recently by G.C. Rossi and G. Veneziano. 
} 
%
%
\section{Introduction}
\label{intro}
\subsection{Measurements of Rapidity Gaps at the Double Diffraction Dissociation}
The recent measurement of diffraction gaps in ATLAS \cite{atlas} has shown that the behavior of their distribution has different character in the different gap ranges. The histogram at the large values of gap indicates some discrete states of gaps. In this area only the process of double diffraction dissociation (DD) gives the contribution. Of course, the discrete pattern can be initiated by poor statistics. Nevertheless, we have to learn here the diagrams that lead to discrete levels of DD gap.
\subsection{Topological Expansion and Pomeron Exchange}
\begin{figure}[htpb]
\centering
  \includegraphics[width=7.0cm, angle=0]{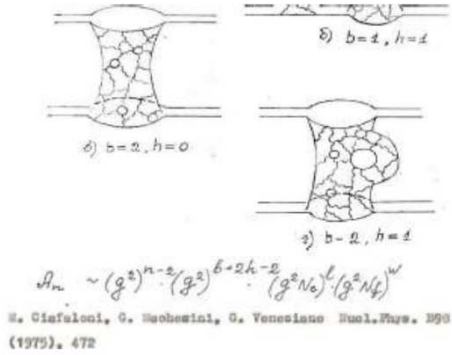}
  \caption{The fragment of graphical presentation of pomeron exchange in the topological expansion, where b is the number of boundaries and h is the number of handles.}
  \label{myphd}
\end{figure}

Pomeron exchange is presented with the topological QCD diagram that is responsible for multi particle production in p-p collisions at LHC energies. The quark-gluon content was drawn in the topological expansion \cite{topologyexp} as the the cylindrical net of gluon exchanges with the random amount of quark-antiquark loops inserted. The topological expansion gives the chance to classify the contributions from general diagrams of multi-particle production in the hadron interactions. This expansion has practically allowed us to develope the Quark-Gluon String Model (QGSM) \cite{kaidalov,qgsm,baryon}. Few orders in topological expansion are graphically presented in the figure from my PhD thesis figure~\ref{myphd}, where the third order is named pomeron with handle. Double diffraction dissociation in this presentation looks like the cylinder of one pomeron exchange with the toroidal handle that left uncut, so no particles have been produced in the central rapidity region.
  
\section{ Double Diffractive Dissociation as an Exchange with Pomeron Torus}

\begin{figure}[htpb]
\centering
  \includegraphics[width=2.0cm, angle=0]{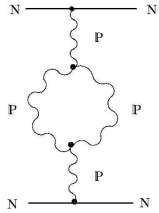}
  \caption{Pomeron loop in the center of one-pomeron exchange.}
  \label{DD}
\end{figure}

Double diffraction dissociation (DD) is a next order in the topological expansion after the pomeron exchange and should be presented as one pomeron diagram with two-pomeron loop in the center, see figure~\ref{DD}. Actually, the DD configuration is similar to cylinder with a handle that takes $(1/9)^2$ from the single pomeron exchange cross section (1.2 percents of $\sigma_{prod}$) at high energies.
If the central pomeron loop was not cut, we are having the DD spectra of produced hadrons: two intervals at the ends of rapidity range, which are populated with multi particle production, and the valuable gap in the center of rapidity.
Looking at two-pomeron loop in this diagram, we are realizing that it is torus in 3D topology. This interesting object should be considered separately in order to reveal some remarkable features for the experimental detection.
 
\section{Baryon Junction-Antijunction Hexagon Net and Discrete Dynamics of DD Gaps}

\begin{figure}[htpb]
\centering
  \includegraphics[width=2.0cm, angle=0]{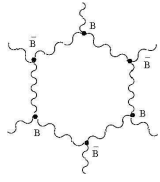}
  \caption{One cell of hexagon net with the SJ and antiSJ.}
  \label{onecell}
\end{figure}

As we remember the pomeron cylinder is built by gluon exchange net, let us consider only three-gluon-connections on the surface of torus (or pomeron loop). This String-Junction type of gluon vertices has been studied in our early researches \cite{baryon,baryonasymmetry} and plays the important role in multiple production of baryons. Since this object brings the baryon charge, the anti-SJ also exists and brings the charge of antibaryon. The only charge-neutral way to construct the net from string-junctions and anti-string-junctions is hexagon where antibaryon charge is following the baryon one as it is shown in the figure~\ref{onecell}. 
 
\begin{figure}[htpb]
\centering
  \includegraphics[width=5.0cm, angle=0]{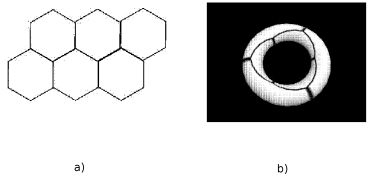}
  \caption{Closed net on the surface of torus: a) six hexagon construction and b) torus covered with six hexagon net.}
  \label{torus}
\end{figure}

The closed net of six hexagons on the torus is shown in the figure~\ref{torus}.

If people are trying to match the eligible number of hexagons, it becomes clear that there is a discrete row: Hexnumb=4, 6, 8, 12, 16, 24, 32, 48, 64 etc, see figure~\ref{torus}. It means that the pomeron torus has certain levels of energy. It leads to discrete gap states at DD \cite{myatlastalk} and to some other signatures in multi-particle production spectra \cite{torusasDD}. The similar construction was presended in \cite{kharzeev} as a complicate fullerene sphere that is build with SJ's. 

\section{More Suggestions on the Pomeron Torus}
It is time to imagine, where to the pomeron torus could contribute. What we have, if our gluon-junction construction that looks like a "compactificated" pomeron string would be released as metastable particle? It is charge neutral QCD cluster with the certain potential energy, which is determined by number of hexagons, Hexnumb. If such cluster would be stable, this is appropriate candidate for the dark matter (DM)\cite{ICPPA}. The masses of these states are suspected similar to the very mass sequence of heavy neutral hadron states invented in \cite{ICHEP}. The reason, why this object is hardly dissipated in the collisions with matter, is following: the atomic numbers of elements in the space are too small in comparison with the number of SJ-antiSJ vertices in our toroidal constructions (let us name them "baryoniums"), therefore compact torus leaves intact after the collision with the less dense light atoms. Since each high energy proton collision in the space, wherever it takes place, contributes 1.2 percents of energy into DM, the valuable DM mass has been accumulated in the Universe, even though some amount of low mass baryoniums dissipates back into baryons and mesons at the collisions with the interstellar light atoms. It seems \cite{ICPPA}, nevertheless, that stable baryonium DM should be mostly concentrated near Supermassive Black Holes due to the huge gravitation pressure. Such a way, DM particles mostly appear in space as the result of  the giant jets radiation and the partial distruction of BHs. This idea has to be verified with more observation of SMBH.

\section{Conclusion}
The topological presentation of pomeron exchange at the proton-proton collision of high energy is cylinder that is covered with quark-gluon net \cite{topologyexp}. I suggest that the process of double diffraction (DD) can be presented as one pomeron exchange with the central loop of two uncut pomeron cylinder loop or torus \cite{qgsm,myPhD}. Taking into account that the junction of three gluons (SJ) has the positive baryon number, as well as the antijunction is of negative baryon charge, our neutral pomeron construction can be covered by only a certain number of hexagons with 3 string junction and 3 antijunction vertices each \cite{torusasDD}.  It is reasonable to expect that the dynamics of rapidity gaps in DD should be determined by the number of hexagons on the surface of pomeron torus. Therefore, the gap distribution in DD events has the discrete structure in the region of large gaps \cite{atlas}. The positive baryon production asymmetries that have been measured at LHC are the demonstration of string junction participation in proton-proton interactions of high energy \cite{baryonasymmetry}. Moreover, the string-junction torus can be released in the course of pp interaction as metastable particle (baryonium) and is getting suspected as "baryonium" Dark Matter candidate \cite{ICPPA}. The possibility of production of the states with many string junctions has been discussed recently by G.C. Rossi and G. Veneziano \cite{newveneziano}. 

\section{Acknowledgments}
Author would like to express her gratitude to Oleg Kancheli for numerous discussions and to Vladimir Tskhay for designing the figure with torus.


\begin{thebibliography}{99}
\bibitem{atlas}ATLAS collab., Eur.Phys.J. C{\bf 72}(2013)1926, e-print - arXiv:1201.2808.
\bibitem{topologyexp}G.C.Rossi  and G. Veneziano, Phys.Lett. B{\bf 70}(1977)255.
\bibitem{kaidalov}A.B. Kaidalov, Phys.Lett. B{\bf 116} (1982)459.
\bibitem{qgsm}Quark-Gluon String Model, A.B. Kaidalov and K.A. Ter-Martirosyan, Sov.J.Nucl.Phys.{\bf 39} (1984)1545, {\it ibid} {\bf 40} (1984)211. 
\bibitem{baryon}A.B.Kaidalov and O.I.Piskunova , Z.Phys.C {\bf 30} (1986) 145.
\bibitem{myPhD}O.I. Piskunova, PhD thesis 1988, in Russian.
\bibitem{myatlastalk}O.I.Piskounova, the talk at ATLAS Diffraction Working Group meeting, CERN, 20 April 2015.
\bibitem{torusasDD}O.I.Piskounova, book of abstracts to Physics Beyond Colliders, CERN, September 2016, e-print - arXiv:1702.02769.
\bibitem{kharzeev}T.Csorgo, M.Gyulassy and D.Kharzeev, in the Proceedings of "International Symposium on Multiparticle Dynamics 2000" WSCI (2001) 616, e-print - arXiv:hep-ph/0102282.
\bibitem{ICHEP}O.I.Piskounova, in the Proceedings of "38th International Conference of High Energy" PoS(ICHEP2016)711, Chicago, USA, August 2016, e-print - arXiv:1602.08003.
\bibitem{baryonasymmetry} O.I. Piskounova , Phys.Atom.Nucl. {\bf 70}(2007) 1107, M.A. Erofeeva and O.I. Piskounova, Nonlin.Phenom.Complex Syst. {\bf 12} (2009) 425, e-print - arXiv:hep-ph/0604157.  
\bibitem{ICPPA}O.I.Piskounova, the talk at ICPPA, Moscow, October 2018, e-print - arXiv:1812.02691.
\bibitem{newveneziano} G.C.Rossi and G. Veneziano, JHEP {\bf 1606} (2016) 041.
\end{thebibliography}
\end{document}